\documentstyle[12pt,aasms4]{article}

\newcommand{\be}{\begin{equation}}
\newcommand{\ee}{\end{equation}}
\newcommand{\etal}{{\it et al.}}

\newcommand{\bef}{\begin{figure}}
\newcommand{\eef}{\end{figure}}
\def\spose#1{\hbox to 0pt{#1\hss}}
\def\ltapprox{\mathrel{\spose{\lower 3pt\hbox{$\mathchar"218$}}
 \raise 2.0pt\hbox{$\mathchar"13C$}}}
\def\gtapprox{\mathrel{\spose{\lower 3pt\hbox{$\mathchar"218$}}
 \raise 2.0pt\hbox{$\mathchar"13E$}}}
\def\inapprox{\mathrel{\spose{\lower 3pt\hbox{$\mathchar"218$}}
 \raise 2.0pt\hbox{$\mathchar"232$}}}

\slugcomment{Subimitted to Astrophys. J. Letters}

\lefthead{Montuori and Sylos Labini }
\righthead{Angular correlations of galaxy distribution}

\begin{document}

\title{Angular correlations of galaxy distribution}

\author{M. Montuori\altaffilmark{1}  and 
F. Sylos-Labini \altaffilmark{1,2}}
\affil{INFM Sezione Roma1,       
	      Dip. di Fisica, Universit\'a "La Sapienza"}

\altaffiltext{2}{ D\'ept.~de Physique Th\'eorique, Universit\'e de Gen\`eve, 
24, Quai E. Ansermet, CH-1211 Gen\`eve, Switzerland}

\begin{abstract}
We study the angular correlations of various galaxy catalogs
(CfA1, SSRS1, Perseus-Pisces, APM Bright Galaxies and Zwicky).
We find that the angular correlation 
exponent is $\gamma_a= 0.1 \pm 0.1$ rather 
than $\gamma_a=0.7$ as usually found by the standard correlation function 
$\omega(\theta)$.  We identify the problem in the artificial 
decay of 
$\omega(\theta)$. Moreover we find that no characteristic angular scale
is present in any of the analyzed catalogs. Finally we show that
 all the available data are consistent with each other and 
the angular distribution of galaxies is quite naturally 
compatible with a fractal
structure with $D \approx 2$.
\end{abstract}


\section{Introduction}

One of the most important elements in the discussion about 
galaxy correlations, is  
 the analysis of angular distributions (see \cite{dav96} and 
\cite{pmsl96} for a review of the problem). Angular catalogs are
qualitatively inferior to  three dimensional  ones because they 
correspond to the angular projection and 
do not contain any information on the third coordinate. However,
the fact that they contain more galaxies than  
the 3-d
catalogs has led some authors to assign an excessive importance 
to these catalogs and  they are supposed to represent a clear 
evidence for homogeneity (\cite{pee93,dav96}). 
Actually the interpretation of
 angular catalogs is quite
 delicate and ambiguous for a variety 
of reasons which are usually 
neglected as we show below (see \cite{cp92} and \cite{slmp97}
for a more detailed discussion of the subject).
It is important to
stress that the existence of large scale structures
of galaxies has been found only in redshift surveys, while 
 angular catalogs are relatively uniform. For this reason the 
reconstruction of 3-d properties of the galaxy distribution
from angular ones, must be  based on a series of assumptions 
which have  be tested in real data. We show that the 
usual hypotheses used so far, contradict the behavior
found in the data analysis of angular correlations, 
free of any a priori assumption.

Usually the analysis of angular correlations of galaxies 
is performed through the two point correlation function 
$\omega(\theta)$. This analysis allows one to determine 
a well defined characteristic scale in the angular distribution
(defined by $\omega(\theta_0) = 1$), and the correlation 
exponent at small angular separation is found to be 
$\gamma_a = 0.7$ (\cite{gp77,mad90,lov96}). This value
of the power law exponent is claimed to be compatible
with the value $\gamma=1.7$ found in  3-d samples,
by the $\xi(r)$ analysis (\cite{pee80,pee93}).

Some years ago we criticized this approach and 
proposed a new one based on the 
{\it concepts and methods of 
modern Statistical Physics} (\cite{pie87,cp92,pmsl96,mont97,slmp97}). 
Usual
statistical methods applied on galaxy catalogs, 
which are based on the assumption of homogeneity, 
 are found to be inconsistent for all the length scales probed so far, and 
a new, more general, conceptual framework is necessary to identify
the real physical properties of galaxy structures.
The 
result of the new analysis of redshift surveys, 
is that galaxy structures are highly irregular and self-similar: 
all the available data are consistent with each other and 
show fractal correlations, with dimension $D \simeq 2$ which 
corresponds to $\gamma \simeq 1$.

In this paper we clarify the conceptual 
problems of the usual $\omega(\theta)$
analysis. First we show that it gives 
a spurious result for the case 
of fractal structures. 
Then we introduce the correct statistical quantity that should 
be used for the characterization of the angular properties of 
irregular (scale-invariant) distributions as well as for 
regular ones.
Moreover, we present the correlation  analysis 
of various angular galaxy catalogs (SSRS1, CfA1, APM Bright Galaxies, 
Perseus-Pisces
and Zwicky). The various catalogs are shown to be consistent
with each other. Our main results are: 
i) The angular correlation exponent is $\gamma_a=0.1 \pm 0.1$ (instead
of $\gamma_a=0.7$ as usually found) ii) No characteristic 
angular scale is present in the available samples. 
These results are  fully  compatible with those 
found  in three 
dimensional catalogs, i.e.
$D \approx 2$ up to the limit of the 
available samples without any tendency towards homogenization (\cite{mont97,slmp97}).
iii) Finally, we comment on the amplitude 
of the angular fluctuations expected for the case of a fractal distribution
with $D \approx 2$.

\section{The Conditional average density and  $\omega(\theta)$ analysis}

The standard method used to analyze angular catalogs,
is based on the assumption that galaxies are correlated only at small
distances. In such a way  the effect of the large spatial inhomogeneities
is not considered at all. Under this  assumption, which is
not
supported by any observational evidence, it is possible to derive
Limber's equation (\cite{lim70,lim71}). In practice,
the angular analysis is performed by computing the two 
point correlation function
\be
\label{eq1}
\omega(\theta) = \frac{\langle n(\theta_0)n(\theta_0+\theta) \rangle}
{\langle n \rangle} -1
\ee
where $\langle n \rangle$ is the average density in the survey.
This function is the analog of $\xi(r)$ for the 3-d analysis.
The results of such an analysis are quite similar to the 
three dimensional ones (\cite{pee80,pee93}). 
In particular, it has been obtained that, 
in the limit of small angles,
\be
\label{eq2}
\omega(\theta) \sim \theta^{-\gamma+1}
\ee
with $\gamma \approx 1.7$ (i.e. $\gamma_a=\gamma-1=0.7$).
 It is possible to show (\cite{pee80}) that,
in the Limber approximation (Eq.\ref{eq2}), the angular correlation
function 
 corresponds to
$\xi(r) \sim r^{-\gamma}$ for its three dimensional counterpart
(in the case $\gamma > 1$).

We now study the case of {\it a self-similar angular distribution}
so that, if such properties are present in real catalogs, we are 
able to recognize them correctly.
Of course, if the distribution is homogenous, we are able to
reproduce the same results obtained by the 
$\omega(\theta)$ analysis.
Hereafter we consider the case of small angles ($\theta \ltapprox 1$), that
is quite reasonable   for the catalogs investigated so far.
In this case the number of points within a cone of opening angle 
$\theta$ scales as 
\be
\label{eq3}
N(\theta)= B_a \theta^{D_a} 
\ee
where $D_a$ is the fractal dimension corresponding to the 
angular projection and $B_a$ is related to the lower cut-off of the 
distribution. Eq.\ref{eq3} holds from every occupied point, and 
in the case of an homogenous distribution we have $D_a=2$.
Following Coleman \& Pietronero (1992),
we define the conditional average density as
\be
\label{eq4}
\Gamma(\theta)= \frac{1}{S(\theta)} \frac{dN (\theta)} {d\theta} = 
\frac{BD_a}{2\pi} \theta^{-\gamma_a}
\ee
where $S(\theta)$ is the differential solid angle element 
($S(\theta) \approx 2 \pi \theta$ for $\theta \ll 1$) 
and $\gamma_a=2-D_a$
is the angular correlation exponent (angular 
codimension). The last equality 
holds in the limit $\theta < 1$. From the very definition of 
$\Gamma(\theta)$ we conclude that
\be
\label{eq5}
\omega(\theta)=\frac{\Gamma(\theta)}{\langle n \rangle} -1 \; .
\ee
A first important consequence of Eq.\ref{eq5} is that 
if $\Gamma(\theta)$ has a power law behavior, and  $\omega(\theta)$ is
a power law minus one. This corresponds to a break in the log-log plot
for angular scales with $\omega(\theta) \ltapprox 1$. We show in 
Fig.1 the behaviour of such a quantity. 
\placefigure{fig1}
The codimension found by fitting $\omega(\theta)$
with a power law function is higher than the real one. This is an important 
effect which
has never been considered before. 
In Fig.1 we show also the $\omega(\theta)$ for the APM Bright Galaxies 
catalog (see below), 
that is fitted quite well by Eq.\ref{eq5} with $D = 1.92$.
Also the 3-d $\xi(r)$ is affected by the same 
problem (\cite{slmp97}). The second important point is that the break
of $\omega(\theta)$ in the log-log plot is clearly artificial and does not
correspond to any characteristic scale of the original distribution.
The basic 
problem is that in the case of a scale-invariant distribution the 
average density in Eq.\ref{eq1} is not well defined, as it depends 
on the sample size (\cite{cp92}).

Before we proceed, it is useful to recall the theorem for 
orthogonal projection of fractal sets.
Orthogonal
 projections preserve the 
sizes of objects. If an object of fractal dimension
$D$, embedded in a  space of dimension $d=3$, is projected on a plane
(of dimension $d'=2$) it is possible to show that 
the projection has dimension $D'$ with (\cite{man82,fal90,cp92})
\be
\label{epro1}
D'=D \; \; \mbox{if} \; \; D<d'=2 \; ;
 \; D'=d' \; \; \mbox{if} \; \; D \ge d'=2 \; .
\ee
This explains, for example, why clouds which have fractal dimension
$D \approx 2.5$, give rise to a compact shadow of dimension $D'=2$.
The angular projection represents a more complex problem 
due to the mix of different 
length scales. Nevertheless 
the theorem given by Eq.\ref{epro1} can be extended 
to the case of angular projections in the limit of 
small angles ($\theta < 1$). Therefore according to 
Eq.(6) we have $D_a=D'$

\section{Angular properties of galaxy catalogs}

We have analyzed the angular properties of the following catalogs:
CfA1 (\cite{huc83}), SSRS1 (\cite{dac91}), Perseus-Pisces (\cite{hg88})
Zwicky {\cite{zwi}) and APM-Bright galaxies (APM- BG - \cite{lov96}).
The results are shown in 
Fig.\ref{fig2}.
\placefigure{fig2}
It turns out 
that all the catalogs show consistent correlation properties.
The angular 
fractal dimension is $D_a \approx 0.1 \pm 0.1$ 
(depending on the sample analyzed). No characteristic 
angular scale is present in any of the analyzed catalogs. 

A point which we want to stress 
is that $\Gamma(\theta)$ has been computed only in circular 
shells. Therefore 
we have limited our analysis to an
effective depth
${\theta_M}$ which  is of the order of the radius of the maximum
circle fully contained in the sample area.
 The reason why
$\Gamma(\theta)$ cannot
be computed for angular separations large than  ${\theta_M}$
is essentially the following.
 When one evaluates the correlation
function beyond ${\theta_M}$ ,
then one makes explicit assumptions on what
lies beyond the sample's boundary. In fact, even in absence of
corrections for selection effects, one
is forced to consider incomplete shells
calculating $\Gamma(\theta)$ for $\theta> {\theta_{M}}$,
thereby implicitly assuming that what one would find 
in the part of the
shell not included in the sample is equal to what is inside.
However such a limitation does not affect the power law behavior 
of $\Gamma(\theta)$ at angular separations $\theta \ltapprox \theta_{M}$.

The {\it maximum depth} of a reliable statistical analysis, 
is limited by the angular dimension  of the sample (as previously discussed), 
while the {\it minimum distance}
depends on the number of points contained in the catalog.
For a Poisson distribution (in a catalog with area $A$) 
the mean average distance between near neighbor 
is of the order $\theta_{min} \sim (A/N)^{\frac{1}{2}}$ ($\theta \ll 1 $).
 Due to its dependence on the sample size, such a 
relation does not give an useful quantity 
in the case of a fractal distribution, like the average 
density, while the meaningful 
measure is the average minimum distance 
between neighboring galaxies $\langle \theta_{min}
 \rangle$, which is related 
to the lower cut-off of the distribution. 
If we measure the conditional density at distances 
$ \theta \ll \langle \theta_{min} \rangle$, 
we are affected 
by a {\it finite size effect}. In fact, due the depletion of points at these 
distances, we underestimate the real conditional 
density, finding a higher value 
for the correlation exponent (and hence a lower value for the fractal 
dimension).
For  $\theta \le \langle \theta_{min}
 \rangle$, we find almost no points and 
the slope is
$\gamma_a\approx -2$ (which corrsponds to $D_a \approx 0$). 
In general, when one measures $\Gamma(\theta)$ at distances that correspond to 
a fraction of $\langle \theta_{min} \rangle$, one finds systematically an higher value for  
$\gamma_a$. 
This   trend is completely spurious and due to the depletion of
points at small  distances. The behavior explained here  
 is clearly shown in Fig.\ref{fig3} for the case of the 
APM Bright Galaxies
catalog.
\placefigure{fig3}

\section{Discussion}
The main results presented in this paper are i) the angular codimension
is 
$\gamma_a=0.1 \pm 0.1$ instead of $\gamma_a =0.7$ as usually found by the 
standard correlation analysis; ii) the angular distribution of galaxies
does show any intrinsic angular scale. The reason why 
most astronomers find the break in the angular correlation function
(or angular power spectrum) in the same position independently on the 
depth of the catalog  is due to the fact that 
the fractal dimension the of angular projection turns out to be 
nearly 2 and hence the dependence of the average density
on the sample depth is very weak (see Eq.(4) where $\gamma_a \approx 0$).

All the catalogs show consistent properties.
 The results 
obtained by the $\omega(\theta)$ analysis are therefore spurious. 
The reason for this disagreement lies in the fact that the standard
analysis is not suitable for the characterization of scale invariant
 structures.

The angular distribution of galaxies turns out to exhibit marginal
scale invariance with angular fractal dimension
$D_a = 0.1 \pm 0.1$. Such a result, in view of the theorem
for orthogonal projections of fractal sets (Eq.(6)) is fully compatible
with the existence of a three dimensional fractal structure with dimension 
$D = 1.9 \pm 0.1$ which we have obtained in the 
analysis of the redshift samples
(\cite{slgmp96,pmsl96,mont97,slmp97}). This result alone is marginally 
compatible 
with an homogenous distribution in real space, because if $D >2$ than
we have $D_a = 2$. It results therefore that the angular analysis 
alone cannot be a {\it strong evidence} in favor of either 
a  homogeneous 
or  a fractal 
distribution in space with dimension 2. However, we stress again that 
the result $\gamma_a=0.7$ is just an artefact due to an inconsistent
 data analysis. We refer to \cite{dp90,cp92,slmp97} for a more
detailed discussion on the properties of the angular correlation function,
while  the problem of the projection of a fractal sets is discussed by \cite{durrer}

It is useful to discuss briefly the angular 
 fluctuations expected in the 
case of fractal dimension $D$. It is possible to show (\cite{pee93})
that the mean square fluctuations of the counts in two field 
of angular size $\theta$, with centers separated 
by angular distances $\Theta  \gg \theta$ is given by
\be
\label{flu}
\langle (N_1 - N_2)^2 \rangle \sim \langle N \rangle^2 
(\theta^{-\gamma_a}-\Theta^{-\gamma_a})
\ee
where $\langle N \rangle$ is the number of points over the whole sky
(it depends on the apparent magnitude limit of the sample).
If the value of the fractal dimension approaches two, then
$\gamma_a \rightarrow 0$ and the angular mean square 
fluctuations $\langle (N_1 - N_2)^2 \rangle \rightarrow 0$.
As we find $\gamma = 0.1 \pm 0.1$, this is compatible with a fractal
distribution in space with $D \approx 2$, and explains the 
uniform distribution of angular maps. A fractal distribution in space 
is characterized by having strong inhomogeneities on all sizes, 
while the angular projection can be quite uniform and isotropic 
if $ D \gtapprox 2$, and  exactly this 
appears to be the case in  the galaxy catalogs (redshift and angular)
available up to now.

\acknowledgments

We are grateful to L. Pietronero, J.-P. Eckmann and A. Gabrielli
for useful discussions and suggestions. This work 
has been partially supported by the Italian Space Agency (ASI).

%
%

\clearpage

\figcaption[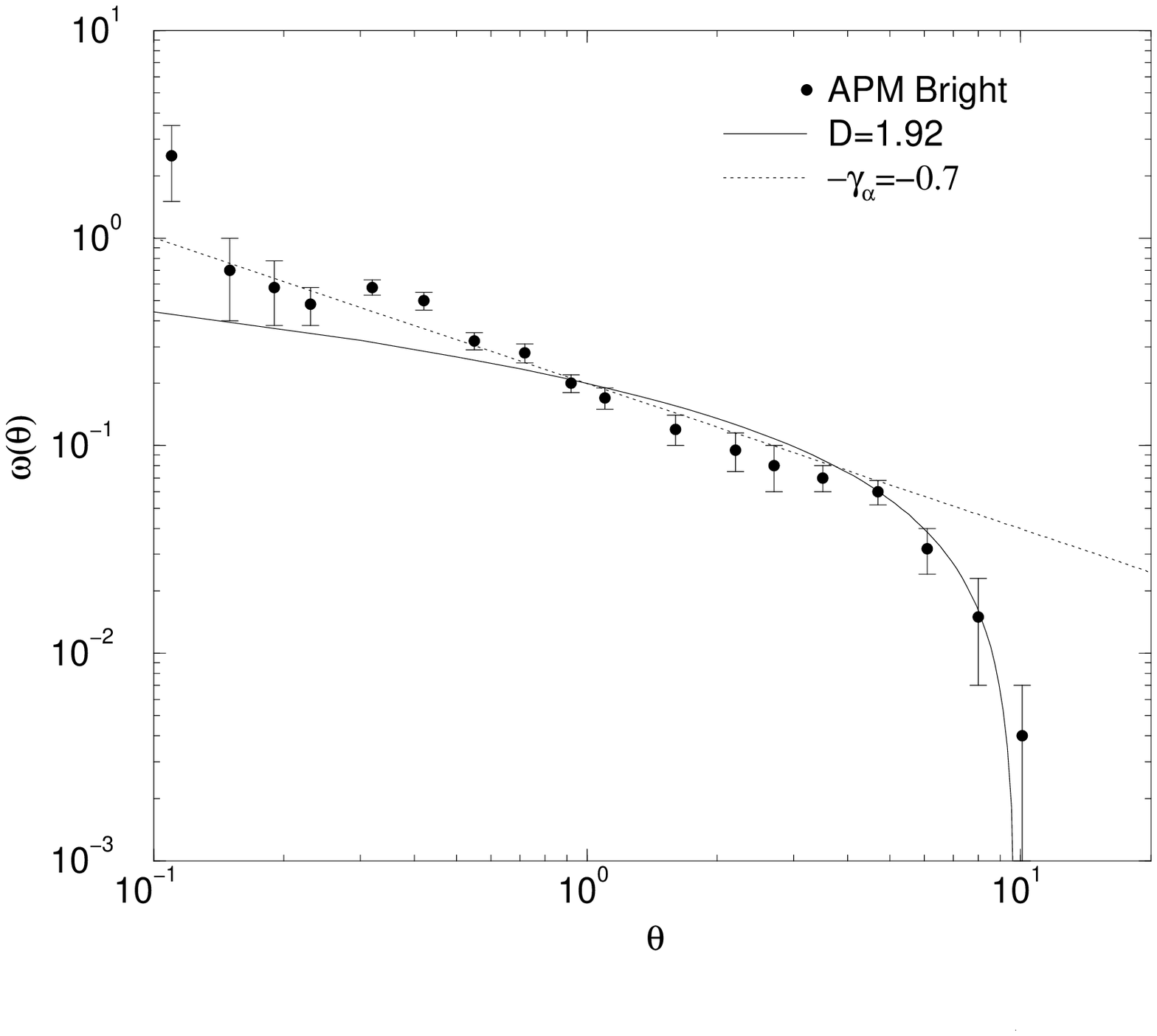]{ In this figure we show the behaviour of
$\omega(\theta)$ (dotted line)
 is the case of a fractal structure with $D_a=1.92$
($\gamma_a=0.08)$ (Eq.5).
It can be seen that the exponent obtained by fitting this function 
with a power law behavior (solid line) is higher than the real one 
($\gamma =-0.7$). Also the break in the power law behaviour is 
completely artificial. The amplitude has been matched to the
one of APM-BG with $m_{lim}=16.44$ (filled circles). \label{fig1}}

\figcaption[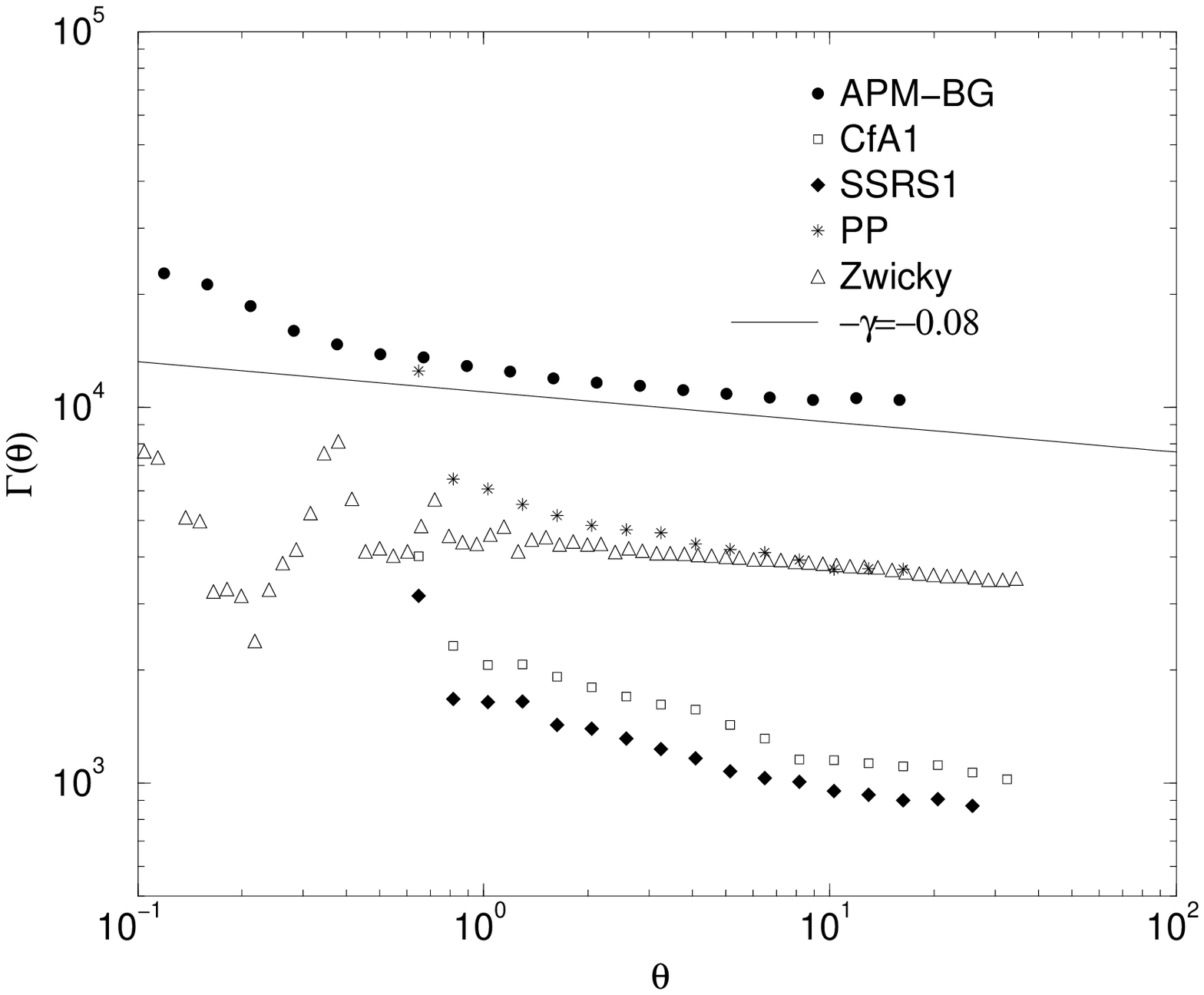] {Angular correlation function $\Gamma(\theta)$
 for the magnitude limited 
samples CfA1, SSRS1, Perseus-Pisces (PP), Zwicky and 
APM Bright galaxies (APM-BG). 
The reference line has a slope $-\gamma_a=-0.08$ 
which corresponds to a fractal dimension $D=1.92$ in the 3-d space. 
The different amplitudes correspond to the different cut in apparent magnitude 
in the various catalogs
\label{fig2}}

\figcaption[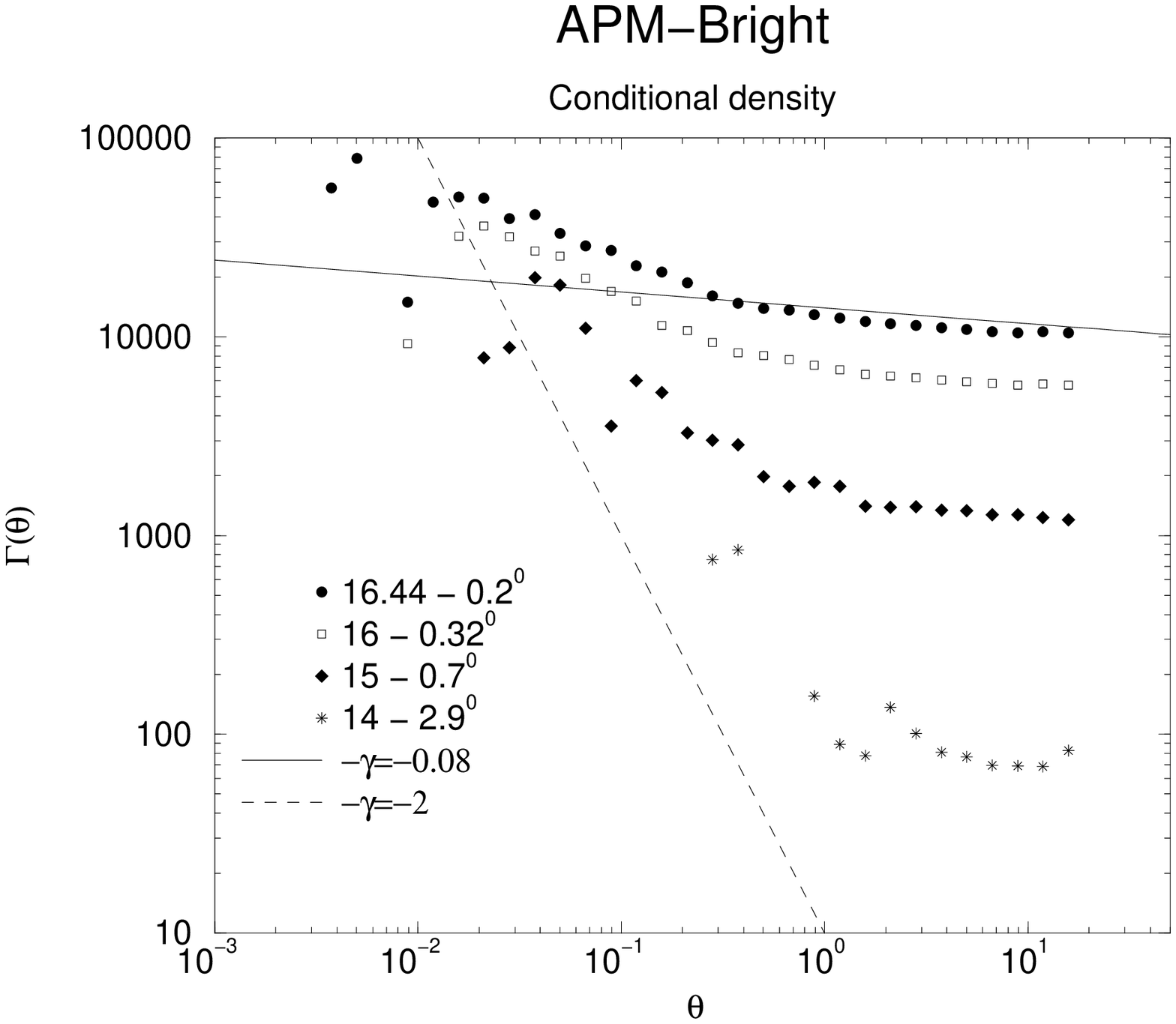] {Angular correlation function $\Gamma(\theta)$
 of the sample APM Bright galaxies (APM-BG), cut at progressively
bright magnitudes. In the figure there are shown the value of the 
apparent magnitude limit and the value of the minimum average angular
distance between neighbor points $\langle \theta_{min} \rangle$.
The solid line has a slope $-\gamma_a=-0.08$ 
which corresponds to a fractal dimension $D=1.92$ in the 3-d space. 
The dotted line has a slope $-\gamma_a=-2$ ($D=0$). The $\theta^{-2}$
behavior is present only at  small angular
 separations $ \theta < 
\langle \theta_{min} \rangle$, and it is due to the depletion 
of points at these scales.
\label{fig3}}


\clearpage

\plotone{ang-fig1.eps}
\clearpage
\plotone{ang-fig2.eps}
\clearpage
\plotone{ang-fig3.eps}

\end{document}